\documentclass[prl,twocolumn,graphicx,amssymb,floatfix]{revtex4}
\begin{document}

\title{Protective measurements of the wave function of a single system}

\author{L. Vaidman}
\affiliation{ Raymond and Beverly Sackler School of Physics and Astronomy\\
 Tel-Aviv University, Tel-Aviv 69978, Israel}

\begin{abstract}
 
My view on the meaning of the quantum wave function and its connection to protective measurements is described. The wave function and only the wave function is the ontology of the quantum theory. Protective measurements  support this view although they do not provide a decisive proof. A brief review of the discovery and the criticism of protective measurements is presented. Protective measurements with postselection are discussed.
    \end{abstract}
\maketitle

\section{Introduction}

In the first graduate course of quantum mechanics I remember asking the question: ``Can we consider the wave function as a description of a single quantum system?'' I got no answer. Twelve years later, in South Carolina, after I completed my Ph.D. studies at Tel Aviv  University under the supervision of  Yakir Aharonov in which we developed the theory of weak measurements \cite{AV90}, I asked Aharonov: Can we use weak measurement to observe the wave function of a single particle?

At that time I already became a strong believer in the many-worlds interpretation (MWI) of  quantum mechanics \cite{SEP} and had no doubt that a single system {\it is} described by the wave function.  Yakir Aharonov never shared with me the belief in the MWI. When we realised that  using what is called now {\em protective measurements}, we can, under certain conditions, observe the wave function of a single quantum system,  he was really excited by the result. At 1992 I was invited to a conference on the Foundations of Quantum Mechanics in Japan where I presented this result: ``The Schr\"{o}dinger wave is observable after all!''\cite{AfterAll}. Then I went  home to Tel Aviv where I finished writing a letter which received mixed reviews in PRL, while Jeeva Anandan, working on the topic with Aharonov in South Carolina,  wrote a paper  accepted in PRA \cite{PRA}. After acceptance of the PRA paper it was hard to fight the referees in PRL, but PLA accepted it immediately \cite{PLA}.

I do not think that protective measurements provide a decisive answer to my question in the graduate school. I came to believe that a single quantum particle is completely described by its Schr\"{o}dinger wave before understanding protective measurements. And  after the publication of our work on protective measurement, many physicists stil view it as an open question. This manifest in the enormous interest to the Pusey,  Barrett, 	and  Rudolph (PBR) paper \cite{PBR} entitled ``On the reality of the quantum state''  which puts strong constrains on the  ensemble interpretation. Also, one of the ``most read'' Nature papers is ``Direct measurement of the quantum wavefunction'' \cite{Lunden}.

The  method of protective measurements provides a plausibility argument: it is more natural to attribute the wave function to a single particle when there is a procedure to observe it.
It  is not a decisive argument because there are some limitations on the measurability of the wave function of a single particle. The quantum state has to be ``protected'', so it is possible to argue that what is measured is not the wave function, but the ``protection procedure''. Still, I can argue that it is  better to say that we measure the quantum state and not its ``protection'' because  there are many different protections for the same wave function, all of  which give rise to the same measurement results.  Moreover, a protection procedure frequently protects many different wave functions, but the protective measurement finds the one which is present.

\section{Why I think that the quantum wave function describes a single quantum system (and everything else)}

I want to believe that Science is capable of explaining everything. We should not explain  every detail since we cannot store all required information.  I think it is enough  to have a theory which can, given an unlimited storage and computational power, explain everything. The theory should agree with experiment. More specifically, any experiment, simple enough to allow theoretically to predict its outcome, should be in agreement with the theory. Classical physics is not such a theory since many experiments like particle interference, atom stability, etc.  contradict classical physics.  Quantum mechanics is such a theory. There is no experiment today contradicting its prediction.  The  ontology of quantum theory is $|\Psi\rangle$ and this is why I believe it is real.

Accepting that there is no ontology beyond the wave function, we admit that every time we perform a preparation procedure of  a particular  quantum state, we end up exactly with the same situation. The failure of Bell inequalities,  the PBR theorem and related results \cite{ColRen,Hardy} support this view. Then, a measurement on an ensemble of identically prepared systems can be viewed as a measurement of the wave function of each system. So, even without protective measurements one can accept the reality of the wave function of a single quantum system.

What might prevent us from considering quantum wave function as real is the collapse of the wave function at quantum measurement. Today there is no a satisfactory physical explanation of the collapse and it is not plausible that such an explanation will be achieved due to nonlocality and randomness of the collapse.

The approach according to which the wave function is not something  real, but  represents a subjective information, explains the collapse at quantum measurement perfectly: it is just a process of updating  the information  the observer has. This approach seems to be highly attractive. The problem is that it has not been successful until now. No one provided an answer to the question: ``Information about what?'' No good candidate for the underlying ontology has been proposed. Bell inequalities  related to quantum measurements performed on entangled particles suggest that such an ontology, if we insist on locality, does not exist. Bell himself, however never gave up  looking for it. He even introduced the concept: local beables \cite{BellBeab}. But in spite of much effort, no attractive theory of beables has been constructed.

One may say that  a successful theory of {\it nonlocal} beables is provided by the de Broglie-Bohm theory, however, it  does not make the wave function epistemic. Bohmians frequently say that ``positions'' represent primary ontology, but the wave function is still an ontology. And in some setups (surrealistic trajectories \cite{sur}, weak measurements \cite{BohmWeak}, protective measurements \cite{PMBohm}) it is the wave function which provides an explanation of the observed results.

Quantum theory  not only corresponds extremely well to our observations, it is also a very elegant theory except for the collapse of the wave function. At 1998 I heard Gottfried saying at a conference in Erice \cite{Got} that: ``The reduction postulate is an ugly scar on what would be a beautiful theory if it could be removed''. I firmly believe it can be removed. There is no experimental evidence for the collapse, only our prejudice that there are no multiple copies of every one of us. Removal of the collapse  leads  to the  MWI. According to the MWI everything is a wave. The Universe is a highly entangled wave. It has natural decomposition into branches corresponding to different worlds in which macroscopic objects are described by well localized wave functions. ``Inside'' a branch every photon emitted from a single photon source  is described by its wave function. The same wave function which everyone will associate with photons emitted by a laser replacing the single photon source.

What are the alternatives? Only a minority remained with a hope to complete the ontology of quantum theory with ``hidden variables''.  A consistent option is to accept that quantum theory is not about what the Nature is, but about what we can say about it, as Bohr preached.  This approach developed into a  popular trend of Qbism \cite{CFS}, a kind of metaphysical nihilism: ``In the quantum world, maximal information is not complete and cannot be completed.'' This nihilism seems unnecessary, because we are given an unprecedentedly successful  deterministic theory of everything, the theory of a quantum wave function.

\section{What is and what is not measurable using protective measurement}

If we are given a single system in an unknown quantum state, there is no way to find out what is this state. It would contradict the no-cloning theorem: if we can find out what is the state, we can prepare many other systems in this state. However, if we are given a single system in a ``protected'' quantum state, we can find out its state.

In general, a single system might not have a description as a pure state. Formally, we can consider a situation when our system is entangled with another system, and the pure state of the composite systems is protected. Then we can observe the density matrix of our system which provides its complete description.  However, if the systems are separate, then the locality of interactions disallows  an efficient protection.

We can specify the wave function of the system by the expectation values of a set of variables. Thus, measuring these expectation values is equivalent to the measurement of the wave function. However, to argue that protective measurements help us  viewing  the wave function of a single system as a ``real'' entity, we need weak  measurements of projections on small regions of space to provide directly the spatial picture of the wave function. Also, weak measurements of local currents \cite{Berry13} allow specifying the phase of the wave function. Of course, only the relative phase can be found, the overall phase is unobservable.

If the wave function has a support in separate regions, the situation is less clear. Local currents do not allow to reconstruct the relative phase between these regions. Another problem is the protection of such a state.
 Clearly, a local classical potential cannot lead to such protection: energy cannot depend on  the relative phase. Although some nonlocal states  can be measured using local interactions with help of measuring devices with entangled parts \cite{AAV86}, many other states cannot be measured in a nondemolition way \cite{PV}.

 A nonlocal state of equal superposition of spatially separated wave packets $\frac{1}{\sqrt 2}(|A\rangle+|B\rangle)$ can be protected using nonlocal measurements. If it is a photon state it can be ``swaped'' to a Bell state of two spins \cite{VHar}. Bell states can be measured in a nondemolition way using measurements of modular sums of spins $(\sigma_z^A+\sigma_z^B){\rm mod}2$, $(\sigma_x^A+\sigma_x^B){\rm mod}2$ \cite{AAV86}. Such measurements require measuring devices with entangled parts. Measurement of a fermion wave function requires also an antiparticle with known phase \cite{AVsingle}. Even if the protection of a particular wave function  is possible, it might not be enough for performing a protective measurement of this wave fucntion.   Nonlocal weak measurements are difficult and current proposals \cite{RS04,BrV} are not efficient.
 The superposition of wave packets of unequal weight, $\alpha|A\rangle+\beta|B\rangle, ~~0<|\alpha|<|\beta|$ cannot be measured in a nondemolition way \cite{PV}, so in this case there is no protection procedure.

 The main motivation for protective measurements was a slogan: ``what is observable is real''. I, however, not fully support it. I do look  for an ontology that can explain all what we observe, but I do not want to constrain ``observation'' to instantaneous nondemolition measurements. I am ready to view the relative phase between two spatially separated parts of the wave function as ``real'',  even if it can be observed only later, when the wave packets overlap.

\section{The  methods of protective measurements and the information gain}

There are two methods for protection of states. First, when it is a nondegenerate eigenstate of some Hamiltonian. Second, based on the quantum Zeno effect, that a frequent measurements of a variable for which the state is a nondegenerate eigenstate are performed. The strength of the protection is characterized by the energy gap to a nearby eigenstate and the frequency of measurements in the Zeno type protection. If the strength of the protection is known, the weakness parameter of weak measurements of the wave function can be calculated. Then, the whole wave function can be found. If the protection strength is not known, we can choose a weakness parameter, make the weak measurement of the wave function and then repeat it. If the result is the same, we know that the measurement is successful. If not, we should ask for another sample. Even if we will need several systems, still, it is not a measurement on a large ensemble.

In the process of protective measurement we gain some information: the wave function is specified by many more parameters than the strength of the protection. There is also some gain of information beyond the information the party which arranges the protection must have. For protection it is enough to know that the state is one of the set of orthogonal states. After the protection and the weak measurement we will know which state of the set is given.

Probably, the Zeno-protection method is easier to understand. First, consider a measurement of the wave function on an ensemble. To know the absolute value of the wave function in a particular point, the projection on a small region of space around this point is measured, and the value is the ratio of the number of times the particle was found there to the number of trials. On every measurement, one of the eigenvalues of the projection operator, 0 or 1, is obtained. The statistical average provides the absolute value of the wave function there.

The projection measurement can be modeled by a von Neumann procedure in which at each trial a pointer, well localized at zero, shifts to 1 or remains at 0. When the uncertainty of the pointer is small, $\Delta \ll 1$, we know with certainty the outcome of each trial. In the corresponding weak measurement, the pointer remains at 0 if the particle is not there, and it is shifted by a small value $\epsilon$, $\epsilon \ll \Delta$, if it is present there. The analysis of weak measurements shows that, up to a very good approximation, after the interaction with  one system, the meter is shifted by $\epsilon |\psi|^2$. If we test now that the system remained in its original state, the probability of the failure is of the order of $\frac{\epsilon^2}{\Delta^2}$. Now, we repeat this procedure with an ensemble of $N=\frac{1}{\epsilon}$ systems using the same measuring device. At the end, the pointer will show the desired value $ |\psi|^2$ with almost unchanged uncertainty $\Delta$. The probability that all the systems will  be found in its original state after the interaction is $(1-\frac{\epsilon^2}{\Delta^2})^{\frac{1}{\epsilon}}$. If we make $\epsilon \ll \Delta^2$, the probability of even one failure goes to zero. It means that we can use the same system every time. This is the Zeno-type protective measurements.

Instead of the test by the projection on the state of interest, we can make  measurements for which  the state we want to measure is one of the eigenstates. If we know that we start with this state, we only need to perform measurement interactions without actually looking at the result. The procedure ensures that we will have our state for the duration of the procedure.

This procedure is different from a measurement on an ensemble not only because we use just one system. Although we have multiple couplings we do not get multiple outcomes and we do not calculate statistical average as in the standard ensemble measurements.

The Zeno-type protection is more of a theoretical construction. Performing frequent verification measurements might be  a very difficult if not impossible task. Another type of protection is to have a Hamiltonian with nondegenerate eigenstates. Experimentally, such a protection is much simpler. We have a system with a nondegenerate ground state. Then we just have to wait long enough time (which can be reduced using some cooling procedure) and the wave function of the ground state is there and protected. It does not require much prior knowledge. Weak measurement coupling after long enough time will provide the information about the wave function. Adiabatic switching of the weak coupling on and off will allow to make the procedure faster.

In recent years, there is a great progress with cooling ions and atoms in traps. Science magazines frequently show pictures of what  looks like a wave function of a trapped particle. But, as far as I understand, these pictures look like a ``wave cloud'' because of the width of the photons which are scattered by the ions. As far as I know, no protective measurements have been performed yet. Recent ``wave function microscopy'' \cite{Cohen} which used a photoionization imaging was also  an ensemble measurement. Although I have no doubts about what will be the outcomes of protective measurements, I think it is of interest to make an effort to perform them, especially since many are reluctant to associate a wave function with a single system.
I hope that some protective measurements will be performed in the near future, maybe along the lines of Nussinov's proposal \cite{Nus}.

The discussion above is about measurement of a stationary wave function. Protective measurements require some period of time to observe the wave function which is supposed to be constant  during this period. Can it shed light on the question of the reality of a nonstationary wave function? If evolution is slow, unitary,  and the protection is strong, then we can observe the wavefunction (with some limited precision) ``in real time''. If we start with the preselected state at the beginning of protection procedure, it will ensure that the wave function will change in time appropriately and the weak coupling of the measurement of the wave will not disturb it significantly. A nonunitary evolution which includes ``collapses'' of the wave function cannot be observed in this way. The weak coupling will not be a problem, but we cannot ensure by any protection procedure that the wave function will collapse  to the right state. This is one more reason not to believe that there is a collapse of a quantum state in Nature.

\section{Protective measurement and postselection}

Protective measurements support the approach according to which the reality of a quantum system is its wave function. Aharonov and myself argued  in many publications that the complete description of a quantum system is a two-state vector consisting of forward and backward evolving wave functions \cite{AV91}. How can these apparently contradicting approaches  peacefully coexist?

In the Zeno-type protective measurements, the backward evolving wave function is identical to the forward evolving wave function. Even if we postselect some other state, the last verification measurement  ``collapses'' the backward evolving state to be identical to the forward evolving one.  It cannot be any other eigenstate of the protection measurement since the measurement specifies both forward and backward evolving states.

For the Hamiltonian type protection the situation is different. We may start, say, with a ground state of a harmonic oscillator and after  performing a protective measurement of the wave function, to  measure the particle position. Whatever outcome we will get, the backward evolving wave function will be very different from the forward evolving wave function. It will not be a stationary state,  but an oscillating wave function.

Protective measurements cause almost no entanglement between the system and the measuring device, so the postselection according to the position of the particle should not change the outcomes of our protective measurements which constituate a picture of the forward evolving  wave function. Weak coupling during a  long period of time of the  projection on every place $x$   shows, at the end, the value of $|\psi(x)|^2$. For all outcomes of the postselection measurement we get the same (approximately correct) result of the density of the wave there. (The postselection which is very far from the center will correspond to some distorted picture since the weak coupling of protective measurements creates some small entanglement.) However, if we will  test whether  during all the interaction time the coupling was to the same value, we will discover that this is not so \cite{AhCo}. To make this test we should perform these protective measurements  on a large pre- and postselected ensemble with all systems originally in the ground state. We should consider a subensemble of particles postselected at localized state at position $x$. In this experiment, we will look at the pointer not at the end, but at various intermediate times. This will allow to find the position of the pointer as a function of time. We will see that the pointer   does not move in the same way during the time of the measurement. Sometimes  it moves fast and sometimes it almost stops. It depends on the backward evolving wave function $\phi(x)$, which changes with time.
The pointer ``feels'' the weak value of the projection operator  $({\rm \bf P}_x)_w=\frac{\phi^*(x)\psi(x)}{\langle \phi|\psi\rangle}$ and it changes with time due to the time dependence of the backward evolving wave function.

This analysis suggests that the Hamiltonian protective measurement does not really measure the density of the wave function, but it measures a time average of a particle density which actually changes with time. To avoid this difficulty we can add a postselection of the state $|\psi\rangle$ at the end of the protective measurement. We know that it will succeed with certainty and then the weak measurement pointer will move with the same velocity during the whole process. Indeed, with the postselection, the weak measurements show weak value of the pre- and postselected system. When the backward wave function is identical to the forward wave function, the weak value is equal to  the expectation value, and it shows the density of the forward evolving wave function.

 In the framework of the MWI,  there is, however, a satisfactory explanation of the situation even without postselection of the original state $|\psi\rangle$. Let us consider again the case of the  postselection measurement of  position. Indeed, in every one out of the different worlds with postselection in different positions, the protective measurement measures time average of the particle density as it is given by the weak value of the projection on a particular location. However, the time averages are subjective to the  observers in worlds with different  postselection positions. These are  not  objective realities. The objective reality is in the universe which incorporates worlds with all possible outcomes of the postselection measurement. In this world, the backward evolving wave function is identical to the forward evolving wave function \cite{Vtime}. To test this we just can look at the preselected ensemble only, without postselection. In this case the pointer will move with constant small velocity ending in the final value $|\psi(x)|^2$.

 In the world with a postselection, subjective reality is best described by the two-state vector. The protective measurement described above showed just forward evolving wave, we observe $|\psi(x)|^2$ and can also observe the local current defined by $ \psi(x ) $. This is because only preselected forward evolving wave was protected. A different  postselected state cannot be orthogonal (such postselection is impossible) and thus it cannot be protected by the same Hamiltonian or by the Zeno measurements. The backward evolving state $ \phi(x) $ changes with time. The effective coupling to the pointer is through weak value which oscillates, but after averaging, it roughly reproduces the expectation value corresponding to the forward evolving wave. If, instead of the protection Hamiltonian of the preselected state we arrange protection by other Hamiltonian (or frequent measurements) which protects the postselected backward evolving state, then our weak coupling of the measurement procedure will yield the backward evolving state  $ \phi(x) $. Forward evolving state will change in time and   will lead to   oscillating weak values averaging to the expectation values corresponding to $ \phi(x ) $.

 At first, it seems that it is impossible to observe the two-state vector on a single system because the same Hamiltonian cannot protect both the forward evolving and a the nonorthogonal backward evolving wave functions. Usually, all Hamitonians are Hermitian so they protect the same set of forward and backward evolving wave functions. However, if we pre- and postselect the whole system including the protection device, then for weak coupling the effective Hamiltonian will be  the weak value of the Hamitonian. Weak value of Hermition operator might be a complex number. Then, the effective Hamiltonians responsible for evolutions of the forward and the backward evolving states are different. Thus, we can, in this way, arrange protection of both forward and backward evolving wave functions even if they are different. This is the protection of the two-state vector \cite{Shimony}.

  The possibility to protect the two-state vector  does not necessarily mean that we can directly observe the two-state vector. The weak coupling to the projection on a particular location $x$ will yield its weak value $({\rm \bf P}_x)_w=\frac{\phi^*(x)\psi(x)}{\langle \phi|\psi\rangle}$. Since weak measurements do not disturb each other significantly, we can measure in parallel weak values of other variables. Protection of the two-state vector will allow then to view the full ``weak measurement reality'' \cite{wmr} specified by the two-state vector. This will allow to reconstruct the two-state vector, but not in a direct way as it was in the protective measurement of the forward (or backward) evolving wave function.

  I have to note that the protective measurement of the two-state vector reality is a highly theoretical concept. Its implementation in laboratory requires highly improbable result of the postselection measurement.

\section{Critique of protective measurements}

There have been many papers criticising our work on protective measurements. According to Paraoanu \cite{Parao}  it has been established that protective measurements program does not work: ``Clearly, to think about the wave function as real, we would have to be able to measure it on a single quantum system. The question of whether this is possible was first raised in the context of
the so-called ``protective'' (weakly disturbing) measurements in the early 1990s, where it was answered in the negative [2].'' He relies on Alter and Yamamoto's  paper \cite{AlterYam} which analyzes the case when the quantum wave function is not protected and, not surprisingly, shows that in this case the wave function  cannot be found \cite{AlYaCom}. Another reference of Paraoanu is ``Impossibility of Measuring the Wave Function of a Single Quantum System'' by D'Ariano and Yuen \cite{DArYu}. All what they claimed is that an unknown and unprotected quantum state cannot be found (due to the no-cloning theorem). Regarding protective measurements they actually say that it works, given that the protection Hamiltonian is known.  They add that known protection Hamiltonian means that its eigenstates are known and that we can find which of the eigenstates is given without breaking the no-cloning theorem and without the need for a weak coupling of protective measurements. They did not distinguish between the party which can find the state and the party that protects the state and do not consider the case that all what is known is  the strength of the protection of the state. Another work Paraoanu cites is a paper with a provocative title by Uffink: ``How to protect the interpretation of the wave function against protective measurements?''\cite{Uff1}. However, Gao recently criticised this paper \cite{Gao} and   Uffink  retracted the main part of his objection \cite{Uff2}.

There are also authors who praised protective measurements. Unruh  \cite{Unru} viewed protective measurements as a demonstration of the reality of certain operators and not of the wave function, but he admitted that ``protective measurements  has broadened our understanding of the quantum measurement process.''
 Ghose and Home \cite{GoHo} in ``An analysis of the Aharonov-Anandan-Vaidman model'' wrote‏: ``the AAV scheme serves to counteract the orthodox belief that quantum mechanics does not say anything empirically meaningful about an individual system.''  Dikson \cite{Dick} considered protective measurement as ``a good reply for the realist'' against empiricist.

 It seems that many of the criticisms follow from misunderstanding triggered by the somewhat misleading examples in our PRA paper \cite{PRA}. One of the examples was a particle in a superposition of two  wave packets which is difficult to observe using protective measurements. The example was technically correct, since there was some tunneling wave connecting the wave packets such that they  were only ``almost'' separate. The Stern-Gerlach experiment had its own difficulties, in particular, related to divergenceless nature of the magnetic field.

Probably the most harsh criticism was a comment by Rovelli \cite{Rov}:
``We argue that the experiment does not provide a way for measuring noncommuting observables without a collapse, does not bear on the issue of the ''reality of the wave function,'' and does not add any particular insight into our understanding (or nonunderstanding) of quantum mechanics.'' I hope,however, that after reading my paper  the following quotation of Rovelli's comment makes his misunderstanding evident: ``To make the problem particularly evident, consider the following experimental arrangement, which is entirely equivalent to the Aharonov-Anandan-Vaidman experiment as far as the interpretation of quantum mechanics is
concerned. First, measure the polarization of a quantum particle. Second, write the outcome of this measurement
on a piece of paper. At this stage assume that we do not know the polarization and we do not know what is written
on the piece of paper. Then make the following protected measurement: Read what is written on the piece
of paper.''

After our clarification \cite{MPM}, Dass  and  Qureshi in  a paper \cite{DaQu} ``Critique of protective measurements''  ``looked at earlier criticisms of the idea, and concluded that most of them are not relevant to the original proposal.'' They argued, however, that ``one can never perform a protective measurement on a single quantum system with absolute certainty. This clearly precludes an ontological status for the wave function.''  I agree that protective measurements do not provide absolute certainty, but as I explained above, this does not prevent me from attributing an ontological status to the wave function.

Protective measurements do not provide a decisive argument for the ontology of the wave function. However, they definitely provide a deep insight into the process of quantum measurement and they strengthen significantly the realist interpretation of the wave function of a single particle. I hope that some protective measurements will be performed in the near future. I believe that they will lead to a significant progress in understanding quantum reality.

This work has been supported in part by  grant number 32/08 of the Binational Science Foundation and the Israel Science Foundation  Grant No. 1125/10.


\begin{thebibliography}{99}

\bibitem{AV90}
Y. Aharonov  and L. Vaidman,
 Properties of a quantum system during the time interval between two measurements,
  Phys. Rev. {\bf A 41}, 11 (1990).


\bibitem{SEP} L.~Vaidman,  Many-Worlds interpretation of quantum
mechanics, {\it Stan. Enc. Phil.},  E. N. Zalta (ed.) (2002),
http://plato.stanford.edu/entries/qm-manyworlds/.

\bibitem{AfterAll}
Y.~Aharonov and L.~Vaidman
The Schrödinger wave is abservable after all!
in {\it Quantum Control and Measurement}, H. Ezawa and Y. Murayama (eds.) 99 (Elsevier Publ., Tokyo, 1993).


\bibitem{PRA}
Y. Aharonov, J. Anandan, and L. Vaidman, Meaning of the wave
function,  Phys. Rev. A {\bf   47}, 4616 (1993).


\bibitem{PLA}
Y. Aharonov and L. Vaidman,
Measurement of the Schrodinger wave of a single particle,
  Phys. Lett. A {\bf  178}, 38 (1993).

\bibitem{PBR}
M. F. Pusey, J. Barrett, and T. Rudolph,
On the reality of the quantum state,
 Nature Phys. {\bf 8}, 476 (2012).

\bibitem{Lunden}
J.S. Lundeen, B. Sutherland, A. Patel, C. Stewart, and C.   Bamber,
 Direct measurement of the quantum wavefunction,
 Nature  {\bf 474},        188    (2011).


\bibitem{ColRen}
R. Colbeck and R. Renner,
Is a System's Wave Function in One-to-One Correspondence with Its Elements of Reality?
Phys. Rev. Lett. {\bf 108},    150402   (2012).

\bibitem{Hardy}
L. Hardy, Are quantum states real? Int. J. Mod. Phys. B, {\bf 27}, 1345012 (2013).


\bibitem{BellBeab}
J.S. Bell,
 Beables for quantum field theory,
 in {J.S. Bell, Speakable and unspeakable in quantum mechanics,} pp. 173–180 (Cambridge: Cambridge University
Press 1987).


\bibitem{sur}
B.G. Englert, M.O.~Scully, G.~S\"{u}ssmann, and H.~Walther,
Surrealistic Bohm trajectories,
  Z. Naturforsch. A {\bf 47}, 1175 (1992).

\bibitem{BohmWeak}
Y. Aharonov and L. Vaidman,
About position measurements which do not show the Bohmian particle position,
in {\it Bohmian Mechanics and Quantum Theory: An Appraisal}, J.T. Cushing, A. Fine and S. Goldstein (eds.), pp. 141-154 (Kluwer, Dordrecht, 1996).

\bibitem{PMBohm}
Y. Aharonov, M. O. Scully, and B.G. Englert,
Protective measurements and Bohm trajectories,
Phys. Lett. A {\bf 263}, 137 (1999).

\bibitem{Got}
K. Gottfried,
 Does quantum mechanics describe the ``collapse'' of the wave function?,
  contribution to the Erice school: {\it 62 Years of Uncertainty} (unpublished) (1989).


\bibitem{CFS}
C. M. Caves, C. A. Fuchs, and R. Schack,
Quantum probabilities as Bayesian probabilities,
Phys. Rev. A {\bf 65}, 022305 (2002).

\bibitem{Berry13}
M.V. Berry,
 Five momenta,
Eur. J. Phys. {\bf 34}, 1337 (2013).

\bibitem{AAV86}
Y. Aharonov, D. Z. Albert, and L. Vaidman,
  Measurement process in relativistic quantum theory,
Phys. Rev. D {\bf 34}, 1805 (1986).

\bibitem{PV}
S. Popescu and L. Vaidman,
Causality constraints on nonlocal quantum measurements,
Phys. Rev. A  {\bf 49}, 4331 (1994).

\bibitem{VHar}
L. Vaidman,
Nonlocality of a single photon revisited again,
Phys. Rev. Lett. {\bf 75}, 2063 (1995).

\bibitem{AVsingle}
Y. Aharonov and L. Vaidman,
Nonlocal aspects of a quantum wave,
Phys. Rev. A  {\bf 61}, 052108 (2000).


\bibitem{RS04}
K.J. Resch and A.M. Steinberg,
Extracting joint weak values with local, single-particle measurements,
Phys. Rev.  Lett. {\bf 92}, 130402 (2004).

\bibitem{BrV}
A. Broduch and L. Vaidman,
 Measurements of non local weak values,
J. Phys.: Conf. Ser. {\bf 174}, 012004 (2009).

\bibitem{Cohen}
S. Cohen, M. M. Harb, A. Ollagnier, F. Robicheaux, M. J. J. Vrakking, T. Barillot, F. Lépine, and C. Bordas,
Wave function microscopy of quasibound atomic states,
Phys. Rev. Lett. {\bf 110}, 183001 (2013).

\bibitem{Nus}
S. Nussinov,
 Scattering experiments for measuring the wave function of a single system,
 Phys. Lett. B {\bf  413}, 382 (1997).


\bibitem{AV91}
Y. Aharonov  and L. Vaidman,
 Complete description of a quantum system at a given time,
J. Phys. A: Math. Gen. {\bf 24}, 2315  (1991).

\bibitem{AhCo}
Y. Aharonov and A. Cohen,
Protective measurement and the Heisenberg representation,
 forthcoming in Gao Shan (ed.) {\it Protective Measurements and Quantum Reality: Toward a New Understanding of Quantum Mechanics,} Cambridge University Press.

\bibitem{Vtime}
L. Vaidman,
Time symmetry and the Many-Worlds Interpretation,
in {\it Many Worlds? Everett, Quantum Theory, and Reality},
S. Saunders, J. Barrett, A. Kent, and D. Wallace eds., (Oxford University Press 2010).


\bibitem{Shimony}
Y. Aharonov, and L. Vaidman,
 Protective measurements of two-state vectors, 
in {\it Potentiality, Entanglement and Passion-at-a-Distance }, R.S.Cohen, M. Horne and J.
Stachel (eds.), BSPS 1-8, (Kluwer, 1997).


\bibitem{wmr}
 L. Vaidman,
Weak-measurement elements of reality,
 Found. Phys. {\bf 26}, 895 (1996).

\bibitem{Parao}
G.S. Paraoanu,
 Extraction of information from a single quantum,
Phys. Rev. A {\bf 83}, 044101 (2011).

\bibitem{AlterYam}
O.Alter and Y. Yamamoto,
Protective measurement of the wave function of a single squeezed harmonic-oscillator state,
Phys. Rev. {\bf  A 53}, R2911 (1996); {\bf  A 56}, 1057 (1997).


\bibitem{AlYaCom}
 Y. Aharonov and L. Vaidman,
  Protective measurement of the wave
function of a single squeezed harmonic-oscillator state -- Comment,
{\it Phys. Rev.  } A {\bf  56}, 1055 (1997).

 \bibitem{DArYu}
 G. M. D'Ariano and H. P. Yuen,
 Impossibility of measuring the wave function of a single quantum system,
 Phys. Rev. Lett. {\bf  76}, 2832 (1996).


\bibitem{Uff1}
J. Uffink,
 How to protect the interpretation of the wave function against protective measurements,
 Phys. Rev. A   {\bf 60}, 3474 (1999).

\bibitem{Gao} 
S. Gao,
 On Uffink's criticism of protective measurements,
  Stud.   Hist. Phil.  Mod. Phys.
{\bf 44},  513 (2013).

 \bibitem{Uff2}
J. Uffink,
Reply to Gao's “On Uffink's criticism of protective measurements”
 Stud.   Hist. Phil.  Mod. Phys.
{\bf 44},  519 (2013).

\bibitem{Unru}
W.G. Unruh,
 Reality and measurement  of the wave function,
 Phys. Rev. A {\bf  50}, 882 (1994).

\bibitem{GoHo}
P. Ghose and D. Home,
An analysis of Aharonov-Anandan-Vaidman
model,  Found. Phys. {\bf 25}, 1105 (1995).

\bibitem{Dick}
M. Dickson,
An empirical reply to empiricism: protective measurement opens the door for quantum realism,
Phil.   Sci.  {\bf 62}, 122 (1995).

\bibitem{Rov}
C. Rovelli,
Meaning of the wave function - Comment,  Phys. Rev. A {\bf  50}, 2788 (1994).



 \bibitem{MPM}
 Y. Aharonov, J. Anandan  and L. Vaidman,
 The meaning of protective measurements,
 Found. Phys. {\bf  26}, 117 (1996).


\bibitem{DaQu}
N.D.H. Dass and T. Qureshi,
 Critique of protective measurements,
Phys. Rev.  A {\bf  59}, 2590 (1999).


\end{thebibliography}
\end{document}